\documentstyle[11pt]{article}
\newcount\Mac  \Mac=0 
\newcommand{\phab}{\phi_{1,2}}
\newcommand{\pha}{\phi_1}
\newcommand{\phb}{\phi_2}

\newcommand{\lx}{\lambda}
\newcommand{\Lx}{\Lambda}

\newcommand{\kx}{\kappa}

\newcommand{\Gammak}{\Gamma_k}

\newcommand{\be}{\begin{equation}}
\newcommand{\ee}{\end{equation}}
\newcommand{\een}{\end{subequations}}
\newcommand{\ben}{\begin{subequations}}
\newcommand{\beq}{\begin{eqnarray}}
\newcommand{\eeq}{\end{eqnarray}}

\def \lta {\mathrel{\vcenter
     {\hbox{$<$}\nointerlineskip\hbox{$\sim$}}}}
\def \gta {\mathrel{\vcenter
     {\hbox{$>$}\nointerlineskip\hbox{$\sim$}}}}

\def\Red{}
\def\Black{}
\def\Blue{}

\newcommand{\lascia}[1]{}
\def\puttag(#1,#2)#3{\put(#1,#2){\makebox(0,0){\rm\Blue #3\Black}}}

\def\circa#1{\,\raise.3ex\hbox{$#1$\kern-.75em\lower1ex\hbox{$\sim$}}\,}

\def\SO{{\rm SO}}

\newcommand{\riga}[1]{\noalign{\hbox{\parbox{\textwidth}{#1}}}\nonumber}

\def\putps(#1,#2)(#3,#4)#5#6{\ifnum\Mac=1 \put(#1,#2){\special{picture #5}}
\else  \put(#3,#4){\includegraphics{#6}} \fi}

\def\Op{{\cal W}}
\def\One{\hbox{1\kern-.24em I}}
\newcommand{\phibounce}{\phi_{\rm b}}

\makeatletter
\def\art{\@ifnextchar[{\eart}{\oart}}
\def\eart[#1]#2#3#4#5#6{{\rm #2}, {\e, #3 \bf #4} {\rm (#6) #5} ({\em #1})}
\def\hepart[#1]#2{{\rm #2, \em#1}}
\newcommand{\oart}[5]{{\rm #1}, {\em #2 \bf #3} {\rm (#5) #4}}
\newcounter{alphaequation}[equation]
\def\thealphaequation{\theequation\hbox to
0.6em{\hfil\alph{alphaequation}\hfil}}
\def\eqnsystem#1{
\def\@eqnnum{{\rm (\thealphaequation)}}
\def\@@eqncr{\let\@tempa\relax \ifcase\@eqcnt \def\@tempa{& & &} \or
  \def\@tempa{& &}\or \def\@tempa{&}\fi\@tempa
  \if@eqnsw\@eqnnum\refstepcounter{alphaequation}\fi
\global\@eqnswtrue\global\@eqcnt=0\cr}
\refstepcounter{equation} \let\@currentlabel\theequation \def\@tempb{#1}
\ifx\@tempb\empty\else\label{#1}\fi
\refstepcounter{alphaequation}
\let\@currentlabel\thealphaequation
\global\@eqnswtrue\global\@eqcnt=0 \tabskip\@centering\let\\=\@eqncr
$$\halign to \displaywidth\bgroup \@eqnsel\hskip\@centering
$\displaystyle\tabskip\z@{##}$&\global\@eqcnt\@ne
\hskip2\arraycolsep\hfil${##}$\hfil& \global\@eqcnt\tw@\hskip2\arraycolsep
$\displaystyle\tabskip\z@{##}$\hfil
\tabskip\@centering&\llap{##}\tabskip\z@\cr}

\def\endeqnsystem{\@@eqncr\egroup$$\global\@ignoretrue} \makeatother

\begin{document}
\begin{quote}
{\em November 1998}\hfill {\bf SNS-PH/98-24}\\
hep-ph/9811438 
\hfill{\bf IFUP-TH/98-54}
\end{quote}
\vspace{1cm}
\begin{center}
{\LARGE\bf\Red Bubble-Nucleation Rates for\\[0.2cm]
Radiatively Induced\\[4mm]
First-Order Phase Transitions}\\[2cm]
\Black\large
{\bf Alessandro Strumia}\\[0.3cm]
\normalsize\em 
Dipartimento di Fisica, Universit\`a di Pisa and\\
INFN, Sezione di Pisa, I-56127 Pisa, Italia\\[0.3cm]
\large {\rm and}\\[3mm] {\bf Nikolaos Tetradis}\\[3mm]
\normalsize\em  Scuola Normale Superiore,\\
Piazza dei Cavalieri 7, I-56126 Pisa, Italia\\[15mm]
\Blue\large\bf Abstract
\end{center}
\begin{quote}\large\indent
We present a consistent calculation of bubble-nucleation rates
in theories of two scalar fields. Our approach is based on the
notion of a coarse-grained free energy that incorporates the
effects of fluctuations with momenta above a given scale $k$.
We establish the reliability of the method 
for a variety of two-scalar models and confirm the conclusions of previous
studies in one-field theories:
Langer's theory of homogeneous nucleation is applicable 
as long as the expansion around the semiclassical
saddle point associated with tunnelling 
is convergent. This expansion breaks down when the exponential
suppression of the rate by the saddle-point action becomes comparable
to the pre-exponential factor associated with fluctuations around the saddle 
point. We reconfirm that Langer's theory is not applicable
to the case of weakly first-oder phase
transitions. We also find that the same is true in general 
for radiatively induced first-order phase transitions.
We discuss the relevance of our results for the electroweak phase
transition and the metastability bound on the Higgs-boson mass.
\end{quote}\Black

\thispagestyle{empty}\newpage\setcounter{page}{1}

\setcounter{equation}{0}
\renewcommand{\theequation}{\thesection.\arabic{equation}}

\section{Introduction}
The calculation of bubble-nucleation rates 
during first-order phase transitions is a difficult problem, both
at the conceptual and at the technical level.
The standard approach to this problem 
is based on the work of Langer~\cite{langer}.
His formalism has been applied to relativistic field theory by
Coleman~\cite{coleman} and Callan~\cite{colcal} and extended by
Affleck~\cite{affleck} and Linde~\cite{linde}
to finite temperature. 
The nucleation rate $I$
gives the probability per unit time and volume to nucleate a certain
region of the stable phase (the true vacuum) within the metastable 
phase (the false vacuum).
Its calculation relies on a semiclassical approximation
around a dominant saddle-point, which is identified with 
the critical bubble. 
This is a static configuration
(usually assumed to be spherically symmetric) within the metastable phase 
whose interior consists of the stable phase.
It has a certain radius that can be determined from the
parameters of the underlying theory. Bubbles slightly larger
than the critical one expand rapidly, thus converting the 
metastable phase into the stable one. 
The nucleation rate is exponentially suppressed by the action
(the free energy rescaled by the temperature) of the critical bubble.
Possible deformations of the critical  bubble
generate a pre-exponential factor.
The leading contribution to this factor  
has the form of a fluctuation determinant and corresponds to the
first-order correction to the semiclassical result. 

For a four-dimensional theory of a real scalar field 
at temperature
$T$, in the limit that thermal fluctuations dominate over
quantum fluctuations, the bubble-nucleation rate
is given by~\cite{colcal}--\cite{linde}
\be
I=\frac{E_0}{2\pi}
\left(\frac{S}{2\pi }\right)^{3/2}\left|
\frac{\det'[\delta^2 \Gamma/\delta\phi^2]_{\phi=\phibounce}}
{\det[\delta^2 \Gamma/\delta\phi^2]_{\phi=0}}\right|^{-1/2}
\exp\left(-S\right). 
\label{rate0} \ee
Here $\Gamma$ is the free energy of the system for a given configuration of the
field $\phi$ that acts as the order parameter of the problem. 
The rescaled
free energy of the critical bubble is $S=\Gamma_b/T
=\left[\Gamma\left(\phibounce(r)\right)-\Gamma(0)\right]/T$,
where $\phibounce(r)$ is the spherically-symmetric
bubble configuration and $\phi = 0$ corresponds to the false vacuum. 
The fluctuation determinants are evaluated either at 
at $\phi = 0$ or around $\phi=\phibounce(r)$. 
The prime in the fluctuation determinant around
the bubble denotes that the three zero eigenvalues 
of the operator $[\delta^2 \Gamma/\delta\phi^2]_{\phi=\phibounce}$
have been removed. 
Their contribution generates the factor 
$\left(S/2\pi \right)^{3/2}$ and the volume factor
that is absorbed in the definition of $I$ (nucleation rate per unit volume). 
The quantity $E_0$ is the square root of
the absolute value of the unique negative eigenvalue.

In field theory, the free energy 
(more precisely the thermodynamic
potential) density $\Gamma$ of a system for homogeneous configurations 
is usually identified with
the temperature-dependent effective potential. This is evaluated 
through some perturbative scheme, such as the loop expansion~\cite{colwein}. 
The profile and the free energy of the bubble are determined
through the potential. This approach, however,
faces three fundamental difficulties:
\begin{itemize}

\item[a)] The effective potential, being the Legendre transform of the
generating functional for the connected Green functions, 
is a convex function of the field. Consequently, it does not 
seem to be the appropriate quantity for the study of tunnelling,
as no structure with more than one minima separated by a barrier
exists.
It has been argued in ref.~\cite{wu} that the appropriate quantity for
the study of tunnelling is the generating functional of the 
1PI Green functions (calculated perturbatively), 
which differs from the effective potential 
in the non-convex regions. However, as we discuss in the following,
the consistent picture must rely on the notion of coarse graining
and on the separation of the high-frequency fluctuations that may be 
responsible
for the non-convexity of the potential, from the low-frequency ones
that are relevant for tunnelling. Such notions cannot be easily 
implemented in the context of perturbation theory.

\item[b)] The fluctuation determinants in the expression for the nucleation
rate have a form completely analogous to the one-loop correction to
the potential. The question of double-counting the effect of
fluctuations (in the potential and the prefactor)
must be properly addressed. 
This point is particularly important 
in the case of 
radiatively induced first-order phase transitions. These are a 
consequence of the appearance of a new vacuum state in the 
theory as a result of the integration of (quantum or thermal) fluctuations
~\cite{colwein}.
A radiatively induced 
first-order phase transition takes place in theories for which the 
tree-level potential has only one minimum, while a second minimum
appears at the level of radiative corrections\footnote{
In ref.~\cite{ewein} an alternative procedure was suggested for the 
treatment of radiatively-induced first-order phase transitions: The
fields whose fluctuations are responsible for the appearance 
of the new vacuum are integrated out first, so that an ``effective''
potential with two minima is generated for the remaining fields. 
Our philosophy is different: We integrate out high-frequency fluctuations
of all fields, so that we obtain an effective low-energy action which we
use for the calculation of the nucleation rate. Our procedure
involves an explicit infrared cutoff in the calculation of the 
low-energy action. This prevents the appearance of non-localities arising
from integrating out massless fields, which may be problematic for
the approach of ref.~\cite{ewein}. For example, the fields that
generate the new vacuum in radiatively-induced first-order phase transitions
are usually massless or very light at the origin of
the potential.
}.

\item[(c)]
Another difficulty 
concerns the ultraviolet divergences that are inherent in
the calculation of the fluctuation determinants 
in the prefactor. An appropriate regularization scheme must be
employed in order to control them~\cite{cott}--\cite{ramos}. 
Moreover, this scheme must be consistent with the one employed
for the absorption of the divergences appearing in the 
calculation of the potential that determines the
free energy of the critical bubble. 
\end{itemize}
In a previous publication~\cite{first},
based on the framework described in refs.~\cite{bubble1,bubble2},
we demonstrated that all the above issues can be resolved
through the implemention of the notion of coarse graining in the 
formalism. As the appropriate quantity for the description of the
physical system we employed
the effective average action $\Gamma_k$~\cite{averact}, which 
is the generalization in the continuum   
of the blockspin action 
of Kadanoff~\cite{kadanoff}. It can be interpreted as a coarse-grained 
free energy at a given scale $k$. 
Fluctuations with 
characteristic momenta $q^2 \gta k^2$ are integrated out
and their effect is incorporated in 
$\Gamma_k$. 
In the limit $k \to 0$,
$\Gamma_k$ becomes equal to the effective action.
The $k$ dependence 
of $\Gamma_k$ is described by an exact flow equation~\cite{exact},
typical of the Wilson approach to the renormalization group~\cite{wilson}. 
This flow equation can be translated into evolution equations
for the invariants appearing in a derivative expansion of
the action~\cite{indices,morris}. In ref.~\cite{first}
we considered only the effective average potential
$U_k$ and a standard kinetic term and neglected higher derivative
terms in the action. We shall employ the same approximation
in this paper also. Its validity is guaranteed by the small anomalous
dimensions in the models we consider. 
The bare theory is defined
at some high scale $\Lx$ that can be identified with the ultraviolet 
cutoff. 
At scales $k$ below the temperature $T$,
the theory can be described in terms of an effective  
three-dimensional action at zero temperature~\cite{trans,me}.
This dimensional reduction indicates the absence of explicit
time dependence for the parameters of the theory at low energy scales.

In ref.~\cite{first} we considered as a starting point the action
$\Gamma_{k_0}$ for a real scalar field
at a scale $k_0$ below the temperature, such that
the theory has an effective three-dimensional description.
We approximated $\Gamma_{k_0}$ by a standard kinetic term and
a potential with two minima. We assumed that this form of the
potential results from the bare potential 
$U_{\Lx}$ after the integration of (quantum and thermal)
fluctuations between the scales
$\Lx$ and $k_0$. 
Some of these fluctuations may correspond to
additional massive degrees of freedom that decoupled above
the scale $k_0$. 
We computed the form of the potential $U_k$ at scales $k\leq k_0$ by
integrating an evolution equation derived from the exact flow 
equation for $\Gamma_k$.  
$U_k$ is non-convex for non-zero $k$, and 
approaches convexity only in the limit $k\to 0$.
The nucleation rate must be computed for $k$ larger than the scale $k_f$
at which the functional integral in the definition of $U_k$ starts 
receiving contributions from field configurations that interpolate between
the two minima. This happens when $-k^2$ becomes approximately equal to 
the negative curvature at the top
of the barrier~\cite{convex}. 
For $k > k_f$ the typical length scale of a thick-wall critical
bubble is $\gta  1/k$. Through the use of $U_k$, 
the first problem 
in the calculation of the nucleation rate 
mentioned above is resolved.

We performed the calculation of the nucleation rate for a range of scales
above and near $k_f$. 
In our approach 
the pre-exponential factor is well-defined and finite,
as an ultraviolet cutoff of order $k$ is implemented in the calculation of
the fluctuation determinants, such that fluctuations
with characteristic momenta $q^2 \gta k^2$ are not included. 
This is a natural consequence of the fact 
that all fluctuations with typical momenta above $k$ are
already incorporated in the form of $U_k$.
This modification also resolves naturally the problem of double-counting 
the effect of the fluctuations.

However, an important issue arises at this point. 
The scale $k$ was introduced in the problem as a mere
calculational tool. If our approach makes sense, the choice
of $k$ should not affect physical 
parameters such as the nucleation rate.
The remarkable outcome of our study was that this
expectation was confirmed. 
We found that the saddle-point configuration 
has an action $S_k$ with a significant $k$ dependence. 
For strongly first-order phase transitions, the nucleation
rate $I = A_k \exp(-S_k)$
is dominated by the exponential suppression.
The main role of the prefactor $A_k$, which is also $k$-dependent,  
is to remove the scale dependence from the total nucleation rate.
The implication of our results 
is that the critical bubble should not be identified
just with the saddle point of the semiclassical approximation.
It is the combination of 
the saddle point and its possible deformations
in the thermal bath (accounted for by the fluctuation 
determinant in the prefactor) that has physical meaning. 
We also found that, 
for progressively more weakly first-order phase transitions,
the difference between 
$S_k$ and $\ln ( A_k/k^4_f )$ diminishes.
This indicates that the effects of fluctuations
become more and more enhanced. 
At the same time, a significant $k$ dependence of the
predicted nucleation rate develops. 
The reason for the above deficiency is clear. 
When the nucleation rate is roughly equal to or smaller 
than the contribution from the prefactor, the effect of 
the next order in the expansion around the saddle point is 
important and can no longer be neglected.
This indicates that there is a limit for the validity of Langer's picture 
of homogeneous nucleation~\cite{langer}. 

\medskip

In this work we describe how our method can be applied
to a more complicated 
system. As such we have chosen a theory of two scalar fields. 
It provides a framework within which we can test
the reliability of our computation of the nucleation rate 
in the case of two fluctuating fields. The 
evolution equation for the potential resembles very closely 
the ones appearing in gauged Higgs theories, with the additional
advantage that the approximations needed in the derivation
of this equation are more transparent. Moreover, we expect
the qualitative conclusions for the region of validity of
Langer's picture of homogeneous nucleation to be valid also 
for gauged Higgs theories. 
The most interesting feature of the two-scalar models is 
the presence of radiatively induced first-order phase transitions.
Such transitions usually take place when the mass of a certain field 
is generated through the expectation value of another. The fluctuations
of the first field can induce the appearance of new minima in the 
potential of the second, resulting in first-order phase transitions
~\cite{colwein}.
As we have already discussed, the problem of double-counting 
the effect of fluctuations is particularly acute 
in such situations. The introduction of a coarse-graining
scale $k$ resolves this problem, by separating the high-frequency 
fluctuations of the system which may be responsible for the 
presence of the second minimum through the Coleman-Weinberg mechanism,
from the low-frequency ones which are relevant for tunnelling. 

In the following sections we present the calculation of the bubble-nucleation
rate for first-order phase transitions in a 
theory of two real scalar fields. 
We do not discuss the evolution of $\Gamma_k$ for $k \gta T$. 
We start the 
evolution at a scale $k_0$ 
sufficiently below the temperature of the system, so that the 
dynamics is three-dimensional to a good approximation\footnote{For readers
who are interested in the details of the
mechanism of dimensional reduction in our approach, detailed
discussions can be found in refs.~\cite{trans,me,twoscalar}
for a variety of models.}. 
As an initial condition we consider a potential $U_{k_0}$, whose
form is determined by the bare potential 
$U_{\Lx}$ and the integration of fluctuations between the scales
$\Lx$ and $k_0$. 
We first establish the reliability of our approach by 
considering potentials $U_{k_0}$ with two minima. Then we turn to
the radiatively induced first-order phase transitions, for which
the second minimum is generated at some point in the evolution 
at a scale $k < k_0$. 
In all cases, we integrate the evolution equation 
for the effective three-dimensional 
theory starting at the scale $k_0$,
and perform the calculation of the nucleation rate
as described earlier in the introduction. 

In the following section we present the model we consider 
and derive the evolution equation for
the potential.
In section~3 we summarize the technical points in the
calculation of the nucleation rate. Our results
are presented in section~4. 
The implications of our results for the first-order phase transitions
in gauged Higgs models (such as the electroweak phase transition)
are given 
in the conclusions of section~5.

\setcounter{equation}{0}
\renewcommand{\theequation}{\thesection.\arabic{equation}}

\section{The model and the evolution equation for the potential}

We consider a model of two real scalar fields $\pha$ and $\phb$. 
The 
effective average action $\Gamma_k(\pha,\phb)$~\cite{averact}
results from the effective integration
of degrees of freedom with characteristic momenta larger than a given 
cutoff $k$. 
This is achieved by 
adding an infrared cutoff term to the bare action,
so that the effective action does not receive contributions from modes with 
characteristic momenta $q^2 \lta k^2$.
In this work we use
the simplest choice of a mass-like cutoff term $\sim k^2 (\pha^2+\phb^2)$,
for which the 
perturbative inverse propagator for massless fields is
$P_k(q) \sim q^2 + k^2$. 
Subsequently,
the generating functional for the connected Green functions is defined,
from which the
generating functional for the 1PI Green functions can be obtained through
a Legendre transformation.
The presence of the modified propagator in the above definitions
results in the effective integration of only the 
fluctuations with $q^2 \gta k^2$. Finally, the 
effective average action is 
obtained by removing the infrared cutoff from the
generating functional for the 1PI Green functions. 

The effective average action $\Gamma_k$ obeys an exact 
flow equation,
which describes its response to variations of the infrared cutoff $k$
~\cite{exact}. This can be turned into evolution equations for the
invariants appearing in a derivative expansion of $\Gamma_k$~\cite{indices}.
In this work we use an approximation which neglects higher derivative 
terms in the action and approximates it by 
\be
\Gammak = 
\int d^3x \left\{ 
\frac{1}{2} \left( \partial^{\mu} \pha~\partial_{\mu} \pha~  
+ \partial^{\mu} \phb~\partial_{\mu} \phb  \right) 
+ U_k(\pha,\phb) \right\}.
\label{twoeleven} \ee
The above action describes the effective three-dimensional theory that 
results from the dimensional reduction of a high-temperature 
four-dimensional theory at scales below the temperature. 
The temperature has been absorbed in a redefinition of the fields and
their potential, so that these have dimensions appropriate
for an effective three-dimensional theory. The correspondence 
between the quantities we use and the ones of the four-dimensional
theory is given by 
\beq
\phab =\frac{[\phi_{1,2}]_4}{\sqrt{T}},\qquad
U(\pha,\phb)=\frac{U_4\left( [\phi_1]_{4},[\phi_2]_{4},T \right)}{T}.
\label{fivethree} \eeq
In this way, the temperature does not appear explicitly in
our expressions. This has the additional advantage of 
permitting the straightforward application of our results
to the problem of quantum tunnelling in a three-dimensional 
theory at zero temperature~\cite{first}. 

The evolution equation for the potential
can be written in the form~\cite{second,twoscalar}
\beq \nonumber
\frac{\partial}{\partial k^2} \left[ U_k(\pha,\phb) - U_k(0,0)\right] &= &
-\frac{1}{8 \pi} \left[
\sqrt{k^2 + M^2_1(\pha,\phb)} 
-\sqrt{k^2 + M^2_1(0,0)}~+ \right.\\ 
&&\phantom{-\frac{1}{8 \pi} }\hspace{-0.3em}\left.+
\sqrt{k^2 + M^2_2(\pha,\phb)}
-\sqrt{k^2 + M^2_2(0,0)} \right],
\label{evpot} \eeq
where $M^2_{1,2}(\pha,\phb) $ are the two eigenvalues of the field-dependent mass matrix,
given by
\be
M^2_{1,2}(\pha,\phb) 
=\frac{1}{2} \left[
U_{11}+U_{22}
\pm  \sqrt{
\left( U_{11}- U_{22} \right)^2
+ 4 U_{12}^2 }
\right],
\label{massm} \ee
with
$U_{i j}\equiv \partial^2U_k/\partial\phi_i\partial\phi_j$.
The only neglected corrections to eq.~(\ref{evpot})
are related to the 
wave-function renormalization of the fields. 
We expect these correction to be small, as the
anomalous dimension is
$\eta \approx 0.035-0.04$ for the models we consider
and the evolution of the potential takes place over
a limited range of $k$. 
We consider models with 
the symmetry $\phb \leftrightarrow - \phb$ throughout
this paper. This means that the expressions for the mass eigenvalues
simplify along the $\pha$-axis:
$M^2_1=\partial^2 U_k/\partial \phi_1^2$, 
$M^2_2=\partial^2 U_k/\partial \phi_2^2$. 
We point out that our method does not 
face the problems of non-convergence of perturbation theory
that often appear in the context of radiative symmetry breaking 
\cite{colwein}. Up to wave-function renormalization effects,
eq.~(\ref{evpot}) is exact and its numerical solution is
very accurate (see below). However, the inclusion of higher-derivative 
terms in the action is required for the quantitative study of 
models with strongly coupled phases, such as the symmetric phase
of the electroweak theory above the critical temperature of the
electroweak phase transition.

\smallskip

The first step of an iterative solution of eq.~(\ref{evpot}) gives
\cite{iterative}
\beq
U_k^{(1)}(\pha,\phb)-U_k^{(1)}(0)=
&~&U_{k_0}(\pha,\phb)-U_{k_0}(0)+
\nonumber \\
&+&
\frac{1}{2}\ln\left[\frac{\det[-\partial^2+k^2+M^2_1(\pha,\phb)]}{
\det[-\partial^2+k^2_0+M^2_1(\pha,\phb)]}
\frac{\det[-\partial^2+k^2_0+M^2_1(0)]}{
\det[-\partial^2+k^2+M^2_1(0)]}\right]
\nonumber \\
&+&
\frac{1}{2}\ln\left[\frac{\det[-\partial^2+k^2+M^2_2(\pha,\phb)]}{
\det[-\partial^2+k^2_0+M^2_2(\pha,\phb)]}
\frac{\det[-\partial^2+k^2_0+M^2_2(0)]}{
\det[-\partial^2+k^2+M^2_2(0)]}\right].
\label{iter} \eeq
For $k\to 0$ this is a regularized one-loop
approximation to the effective
potential. Due to the ratio of determinants, only
momentum modes with $k^2\lta q^2 \lta k_0^2$ are effectively included
in the momentum integrals in eq.~(\ref{iter}).
The above expression demonstrates the form of ultraviolet regularization
of fluctuation determinants that is consistent with the cutoff 
procedure that we described in the beginning of this section in 
the derivation of the evolution equation for the potential. An analogous
regularization will be used in the following section for the fluctuation
determinants in the expression for the nucleation rate.
The solution of eq.~(\ref{iter}) also demonstrates that the decoupling 
of heavy modes is automatically built in the evolution equation for the 
potential. If $M^2_i(\pha,\phb) \gg k_0^2\geq k^2$, the contribution
of the respective mode to this solution is negligible.

We can determine the symmetries of the theory by specifying the form of
the potential at the scale $k_0 < T$ at which we 
start the evolution. As we have already mentioned, we
assume the symmetry 
$\phb \leftrightarrow - \phb$ throughout this paper. 
In section 4 we consider an example of a theory without any 
additional symmetry for the $\pha$ field. However, 
for most of our results and especially for those concerning 
radiatively induced first-order phase transitions, we 
use potentials invariant under 
$\pha \leftrightarrow - \pha$, 
$\phb \leftrightarrow - \phb$, 
$\pha \leftrightarrow \phb$. 
The solutions of eq.~(\ref{evpot}) preserve the above 
symmetries for $k < k_0$. 

For the numerical integration of eq.~(\ref{evpot}), we use the
algorithms described in ref.~\cite{num}, with an appropriate generalization
to the two-field case. As we shall be interested in the form of the
potential along the $\pha$-axis, we perform the integration of the 
evolution equation for $U_k(\pha,\phb)$
in a thin region around this axis. In certain
cases, which would require excessive computer time, we solve 
approximate evolution equations for $U_k$ and $\partial^2 U_k/
\partial \phb^2$ along the $\pha$-axis ($\phb=0$). We control the accuracy
of these approximate solutions by comparing them against the 
full solution for $U_k(\pha,\phb)$ when this is possible. Moreover, we
check our numerical results against exact analytical solutions for 
simplified forms of the evolution equation~\cite{analytical}.
We estimate that the numerical uncertainties introduce an 
error of order 1\% to our results, smaller than the error generated
by the omission of the anomalous dimension.

\setcounter{equation}{0}
\renewcommand{\theequation}{\thesection.\arabic{equation}}

\section{The calculation of the bubble-nucleation rate}

The calculation of the nucleation rate proceeds in complete analogy
to the one described in detail in ref.~\cite{first} for the one-field case.
For this reason, we summarize here only the main steps of the calculation. 
We consider potentials $U_k(\pha,\phb)$ with two minima along the 
$\pha$-axis $(\phb=0)$:
the stable (true)
minimum is located at $\pha=\phi_t$ and the unstable (false) one at
$\pha=\phi_f=0$.

The nucleation rate is exponentially suppressed by 
the action $S_k$ (the rescaled free energy) of the saddle-point configuration
$\phibounce(r)$ that is associated with tunnelling. This is an 
$\SO(3)$-invariant solution of
the classical equations of motion along the $\pha$-axis ($\phb=0$)
which interpolates between 
the local maxima of the potential
$-U_k(\pha) \equiv -U_k(\pha,\phb=0)$. It satisfies the equation 
\be
{d^2\phibounce\over dr^2}+\frac{2}{r}~{d\phibounce\over dr}=
\frac{\partial U_k(\phibounce)}{\partial \pha}, 
\label{eom} \ee
with the boundary conditions 
$\phibounce\rightarrow 0$ for $r \rightarrow \infty$ and
$d\phibounce/dr= 0$ for $r =0$.
The action $S_k$ of the saddle point is given by 
\be
S_k=4 \pi
\int_0^\infty 
\left[ \frac{1}{2}
\left(\frac{d\phibounce(r)}{dr}\right)^2
+U_k(\phibounce(r)) -U_k(0) \right]
r^{2}\,dr.
\label{action} \ee 
The profile of the saddle point can be easily computed 
with the ``shooting'' method~\cite{shooting}. We integrate eq.~(\ref{eom}) numerically,
starting at $r=0$ with a value of $\pha$ near
the true minimum $\phi_t$ and $d\pha/dr=0.$ We then adjust the initial value
of $\pha$ so that the boundary condition 
$\phibounce\rightarrow 0$ for $r \rightarrow \infty$ is satisfied. 

In all the models we consider, 
the minima of the potential and the saddle-point configuration
are located along the $\pha$-axis ($\phb=0$).
The unrenormalized decay rate per unit volume 
from the false minimum towards the true one is then given by 
\cite{coleman,colcal,cott}
\be
I=\frac{E_0}{2\pi}
\left(\frac{S_k}{2\pi}\right)^{3/2}\left|
\frac{\det'[\delta^2 \Gamma_k/\delta\pha^2]_{\pha=\phibounce}}
{\det[\delta^2 \Gamma_k/\delta\pha^2]_{\pha=0}}~
\frac{\det[\delta^2 \Gamma_k/\delta\phb^2]_{\pha=\phibounce}}
{\det[\delta^2 \Gamma_k/\delta\phb^2]_{\pha=0}}
\right|^{-1/2}
\exp({-S_k}). 
\label{rate} \ee
This is analogous to eq.~(\ref{rate0}) after the absorption 
of the explicit factors of $T$ in the redefinition of the fields
and potential and the introduction of a coarse-graining scale. 
If there are several equivalent true vacua, the above rate must be 
multiplied by an appropriate factor, in order to take into account
the possibility of the false vacuum decaying into any of them.

The pre-exponential factor 
corresponds to the first correction to the semiclassical approximation
in the saddle-point method. 
We are considering models with 
the symmetry $\phb \leftrightarrow - \phb$ throughout
this paper. This guarantees that the saddle-point configuration is
located along the $\pha$-axis, where the mass eigenvalues
simplify:
$M^2_1=\partial^2 U_k/\partial \phi_1^2 \equiv U_{11}$, 
$M^2_2=\partial^2 U_k/\partial \phi_2^2 \equiv U_{22}$. 
The numerators in eq.~(\ref{rate}) are the 
fluctuation determinants 
around the saddle-point
\beq
{\det}'\left[ \delta^2 \Gamma_k/\delta\pha^2 \right]_{\pha=\phibounce} &=&
{\det}'\left[-\partial^2+U_{11}(\pha=\phibounce(r))
\right],
\nonumber \\
{\det}\left[ \delta^2 \Gamma_k/\delta\phb^2 \right]_{\pha=\phibounce} &=&
{\det}\left[-\partial^2+U_{22}(\pha=\phibounce(r))
\right],
\label{fluct1} \eeq
while the denominators are
the fluctuation determinants
around the false vacuum $\pha=\phb=0$
\beq
\det\left[ \delta^2 \Gamma_k/\delta\pha^2 \right]_{\pha=0} &=&
\det\left[ -\partial^2+ U_{11}(\pha=0)\right],
\nonumber \\
\det\left[ \delta^2 \Gamma_k/\delta\phb^2 \right]_{\pha=0} &=&
\det\left[ -\partial^2+ U_{22}(\pha=0)\right].
\label{fluct2} \eeq
The differential operator $-\partial^2+U_{11}
(\phibounce(r))$ has three 
zero modes (the three spatial translations of the critical bubble).
The prime over the determinant indicates that 
these modes have to be omitted 
in its calculation. 
Their contribution generates the factor $(S_k/2\pi)^{3/2}$ in 
eq.~(\ref{rate}) and the volume factor
that is absorbed in the definition of $I$ (nucleation rate per unit volume).
The quantity $E_0$ is the square root of
the absolute value of the unique negative eigenvalue of the above operator.
The operator $-\partial^2+U_{22}
(\phibounce(r))$ has only positive eigenvalues, because the bubble 
is stable in the $\phb$ direction at $\phb=0$.

The pre-exponential factor defined in eq.~(\ref{rate})
is in general ultraviolet-divergent and an appropriate regularization
scheme must be employed. 
Within our approach, the form of the regularization 
is dictated by the discussion at the end of the previous section.
The effect of the high-frequency modes has been incorporated in the 
form of the coarse-grained potential $U_k$, which is obtained
through the integration of the evolution equation (\ref{evpot}). 
Fluctuation
determinants computed within the low-energy theory must be 
replaced by a ratio of determinants, in complete analogy to 
eq.~(\ref{iter}). This implies that 
the nucleation rate is given by 
\beq
I&=&A_{1k} A_{2k} \exp({-S_k})
\nonumber \\
\riga{where}\\[-2mm]
A_{1k}&=& \frac{E_0}{2\pi}\left(\frac{S_k}{2\pi}\right)^{3/2}
\left|
\frac{\det'\left[-\partial^2+U_{11}(\phibounce(r)) \right]}
{\det \left[ -\partial^2+k^2 +U_{11}(\phibounce(r))\right]}
~\frac{\det\left[-\partial^2+k^2+U_{11}(0) \right]}
{\det\left[-\partial^2+U_{11}(0)\right]}
\right|^{-1/2},
\nonumber \\
A_{2k}&=& \left|
\frac{\det\left[-\partial^2+U_{22}(\phibounce(r)) \right]}
{\det \left[ -\partial^2+k^2 +U_{22}(\phibounce(r))\right]}
~\frac{\det\left[-\partial^2+k^2+U_{22}(0) \right]}
{\det\left[-\partial^2+U_{22}(0)\right]}
\right|^{-1/2}.
\label{rrate} \eeq
The above form of the pre-exponential factors guarantees that only 
modes with characteristic momenta $q^2 \lta k^2$ contribute to the
nucleation rate. Another feature of eq.~(\ref{rrate}) is
the decoupling of heavy modes with 
$M^2_i \gg k^2$.

The differential operators that appear in eq.~(\ref{rrate}) 
have the general form
\begin{eqnsystem}{sys:W}
\Op_{i \kappa\alpha}&=&-\partial^2 +m_{i\kappa}^2+\alpha W_{ik}(r) 
\label{op} \\
\riga{where}\\[-2mm]
m_{i\kappa}^2&\equiv&U_{ii}(0)+\kappa k^2,\\
W_{ik}(r)&\equiv&U_{ii}(\phibounce(r))-U_{ii}(0),\label{ak}
\end{eqnsystem}
with $i=1$ or 2 and $\kx,\alpha=0$ or 1.
Since the $\Op_{i\kappa\alpha}$ operators are SO(3) symmetric, it is convenient
to use spherical coordinates 
and express the eigenfunctions in terms of spherical harmonics.
This leads to 
\beq
\det \Op_{i\kappa\alpha}
&=&\prod_{\ell=0}^\infty (\det \Op_{i\ell\kappa\alpha})^{2\ell+1},
\nonumber \\
\Op_{i\ell\kappa\alpha}&=&-\nabla^2_\ell+m_{i\kappa}^2+\alpha W_{ik}(r),
\label{opl} \eeq
where
\be
\nabla^2_\ell 
\equiv 
\frac{d^2}{dr^2}-\frac{\ell(\ell+1)}{r^2}
\label{sph} \ee
and $\ell$ is the usual angular quantum number.

The computation of such complicated determinants is made possible by 
a theorem~\cite{erice,cott} that relates ratios
of determinants to solutions of ordinary differential equations.
In particular, we have  
\be
g_{i\ell\kappa}\equiv
\frac{\det \Op_{i\ell \kappa 1}}{\det \Op_{i\ell \kappa 0}}
=\frac{\det[-\nabla^2_\ell+m_{i\kappa}^2+1\cdot W_{ik}(r)]}
{\det[-\nabla^2_\ell+m_{i\kappa}^2+0\cdot W_{ik}(r)]}=
\frac{y_{i\ell\kappa1}(r\to \infty)}{y_{i\ell\kappa0}(r \to \infty)},
\label{theorem} \ee
where $y_{i\ell\kappa\alpha}(r)$ is the solution of the differential equation
\be
\left[-\frac{d^2}{dr^2}+\frac{\ell(\ell+1)}{r^2}
+m_{i\kappa}^2+\alpha W_{ik}(r)\right]y_{i\ell\kappa\alpha}(r)=0,
\label{diffeq} \ee
with the behaviour
$y_{i\ell\kappa\alpha}(r)\propto r^{\ell+1}$ for $r\to 0$.
Such equations can be easily solved numerically with Mathematica~\cite{Mathematica}.
The final expression for the nucleation rate, 
appropriate for an efficient numerical computation~\cite{first}, is
\beq
I &=& \frac{1}{2 \pi}
\left(\frac{S_k}{2\pi}\right)^{3/2}\exp\left(-S_k\right)
\prod_{i=1}^2
\prod_{\ell=0}^\infty c_{i\ell},
\nonumber \\
c_{10} &=& \left( \frac{E_0^2 g_{101}}{\left| g_{100} \right|}
\right)^{1/2},~~~~~~~~~~
c_{11} = \left({g_{111}\over g'_{110}}\right)^{3/2},~~~~~~~~~~
c_{1\ell} = \left({g_{1\ell 1}\over g_{1\ell 0}}\right)^{(2\ell+1)/2},
\nonumber \\
c_{2\ell} &=& \left({g_{2\ell 1}\over g_{2\ell 0}}\right)^{(2\ell+1)/2}.
\label{fin} \eeq
The calculation of $c_{11}$ is slightly complicated because of the necessity to
eliminate the zero eigenvalues in $g'_{110}$. 
Also the (unique) negative eigenvalue $-E_0^2$ 
of $\Op_{1001}$ must be computed for the determination of $c_{10}$. 
How these steps are achieved is described in ref.~\cite{first}, where it
is also checked that the product of an infinite number of terms 
in eq.~(\ref{fin}) is finite. This can be shown by employing 
first-order perturbation theory in $W_{ik}$ for large $\ell$, which gives
\beq
c_{i\ell} \to 1 + \frac{D_i}{\ell^2}+ {\cal O}(\ell^{-4}) 
,\qquad\hbox{with}\qquad
D_i = -\frac{1}{4} k^2 \int_0^\infty 
r^3\, W_{ik}(r)\,dr.
\label{asympt} \eeq
As eq.~(\ref{asympt}) gives a reasonable approximation of $c_{i\ell}$
even for small values of $\ell$ in all the cases we consider, 
we can derive an approximate expression for
the prefactors
\be
\ln( A_{ik}) \approx {\rm sign}(D_i) \sqrt{\left|D_i \right|}\, \pi.
\label{apprpref} \ee
We have checked that the above expression gives a good analytical
approximation to
the numerical results we present in the next section.

\begin{figure}[t]
\begin{center}\hspace{-5mm}
\begin{picture}(16,11)
\putps(0,-0.4)(0,-0.3){fcubic}{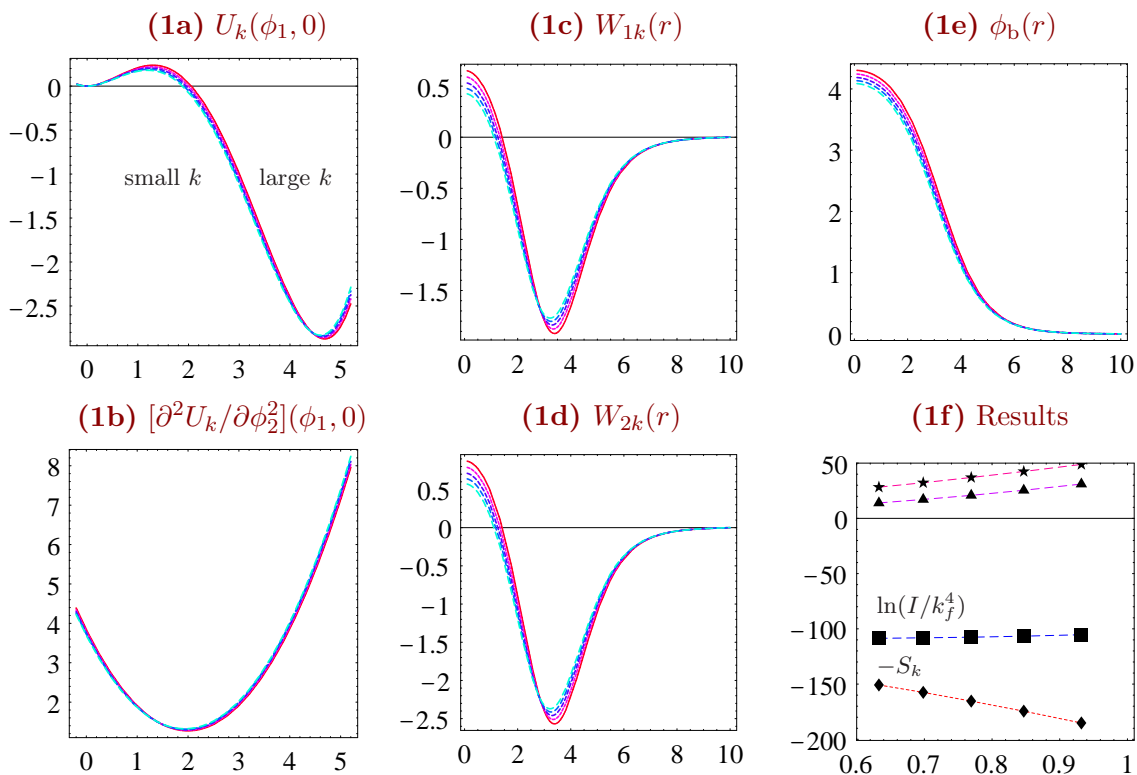}\Red
\put(3.5,10){\makebox(0,0){{\bf (\ref{fig:cubic}a)} $U_k(\phi_1,0)$}}
\put(8.4,10){\makebox(0,0){{\bf (\ref{fig:cubic}c)} $W_{1k}(r)$}}
\put(13.5,10){\makebox(0,0){{\bf (\ref{fig:cubic}e)} $\phi_{\rm b}(r)$}}
\put(3.3,4.8){\makebox(0,0){{\bf (\ref{fig:cubic}b)} $[\partial^2 U_k/\partial\phi_2^2](\phi_1,0)$}}
\put(8.4,4.8){\makebox(0,0){{\bf (\ref{fig:cubic}d) $W_{2k}(r)$} }}
\put(13.5,4.8){\makebox(0,0){{\bf (\ref{fig:cubic}f)} Results}}
\Black
\put(2,7.9){\footnotesize small $k\qquad$ large $k$}
\put(12,1.4){\footnotesize $-S_k$}
\put(12,2.2){\footnotesize $\ln(I/k_f^4)$}
\end{picture}
\caption[SP]{\em The steps in the computation of the nucleation rate for a model with
initial potential given by eq.~(\ref{pot1}) with $m_{2k_0}^2=-m_{1k_0}^2
=0.1~k_0^2$, 
$\lambda_{k_0}=g_{k_0}=0.1~k_0$, $J_{k_0}=0.6~k_0^{5/2}$.
The calculation is performed between the scales $k_i=e^{-0.8} k_0$ and $k_f=e^{-1.2} k_0$.
All dimensionful quantities are given in units of $k_f$.
In fig.~\ref{fig:cubic}f we plot the saddle-point action (diamonds),
the two prefactors $\ln ( A_{1k}/k^4_f )$ (stars) and $\ln ( A_{2k} )$ (triangles),
and the nucleation rate $\ln(I/k_f^4)$ (squares)
as a function of $k/\sqrt{U_{11}(\phi_t,0)}$.
\label{fig:cubic}}
\end{center}\end{figure}

\begin{figure}[t]
\begin{center}\hspace{-5mm}
\begin{picture}(16,11)
\putps(0,-0.4)(0,-0.4){fmedg}{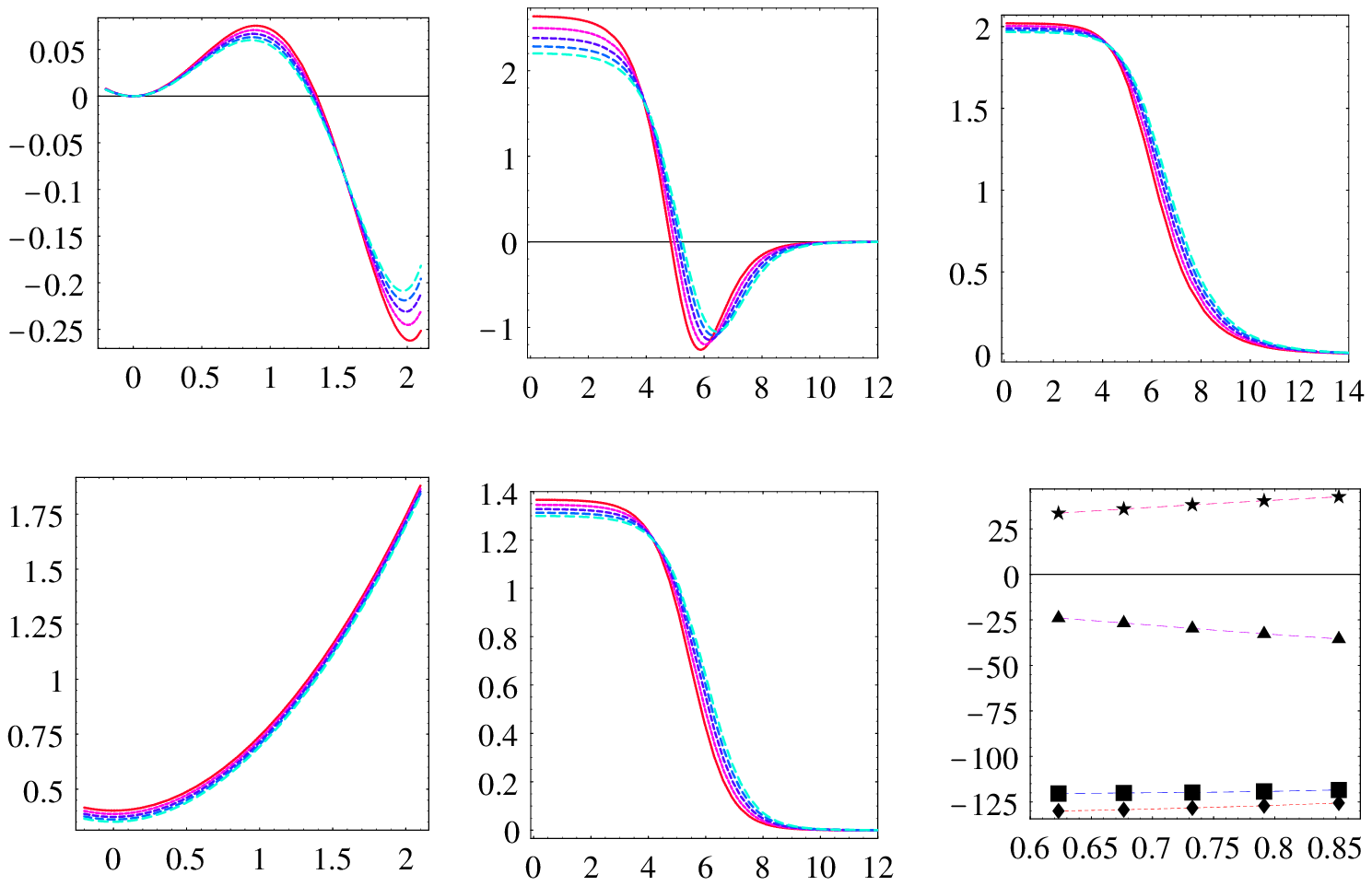}\Red
\put(3.5,10){\makebox(0,0){{\bf (\ref{fig:medg}a)} $U_k(\phi_1,0)$}}
\put(8.4,10){\makebox(0,0){{\bf (\ref{fig:medg}c)} $W_{1k}(r)$}}
\put(13.5,10){\makebox(0,0){{\bf (\ref{fig:medg}e)} $\phi_{\rm b}(r)$}}
\put(3.3,4.8){\makebox(0,0){{\bf (\ref{fig:medg}b)} $[\partial^2 U_k/\partial\phi_2^2](\phi_1,0)$}}
\put(8.4,4.8){\makebox(0,0){{\bf (\ref{fig:medg}d) $W_{2k}(r)$} }}
\put(13.5,4.8){\makebox(0,0){{\bf (\ref{fig:medg}f)} Results}}
\Black
\put(2.5,8.7){\footnotesize small $k$~~~large $k$}
\put(12,1.3){\footnotesize $\ln(I/k_f^4)$}
\put(12,2.3){\footnotesize $\ln (A_{2k})$}
\put(12,3.6){\footnotesize $\ln (A_{1k}/k_f^4)$}
\end{picture}
\caption[SP]{\em As in fig.~\ref{fig:cubic}, but for a model with
initial potential given by eq.~(\ref{pot2})
with $\phi^2_{0k_0}=2~k_0$, $\lx_{k_0}=0.4~k_0$, $g_{k_0}=0.3~k_0$ and $\nu_{k_0}=1$.
The calculation is performed between the scales $k_i=e^{-0.4} k_0$ and $k_f=e^{-0.8} k_0$.
All dimensionful quantities are given in units of $k_f$.
\label{fig:medg}}
\end{center}\end{figure}

\begin{figure}[t]
\begin{center}\hspace{-5mm}
\begin{picture}(16,11)
\putps(0,-0.4)(0,-0.4){fmaxg}{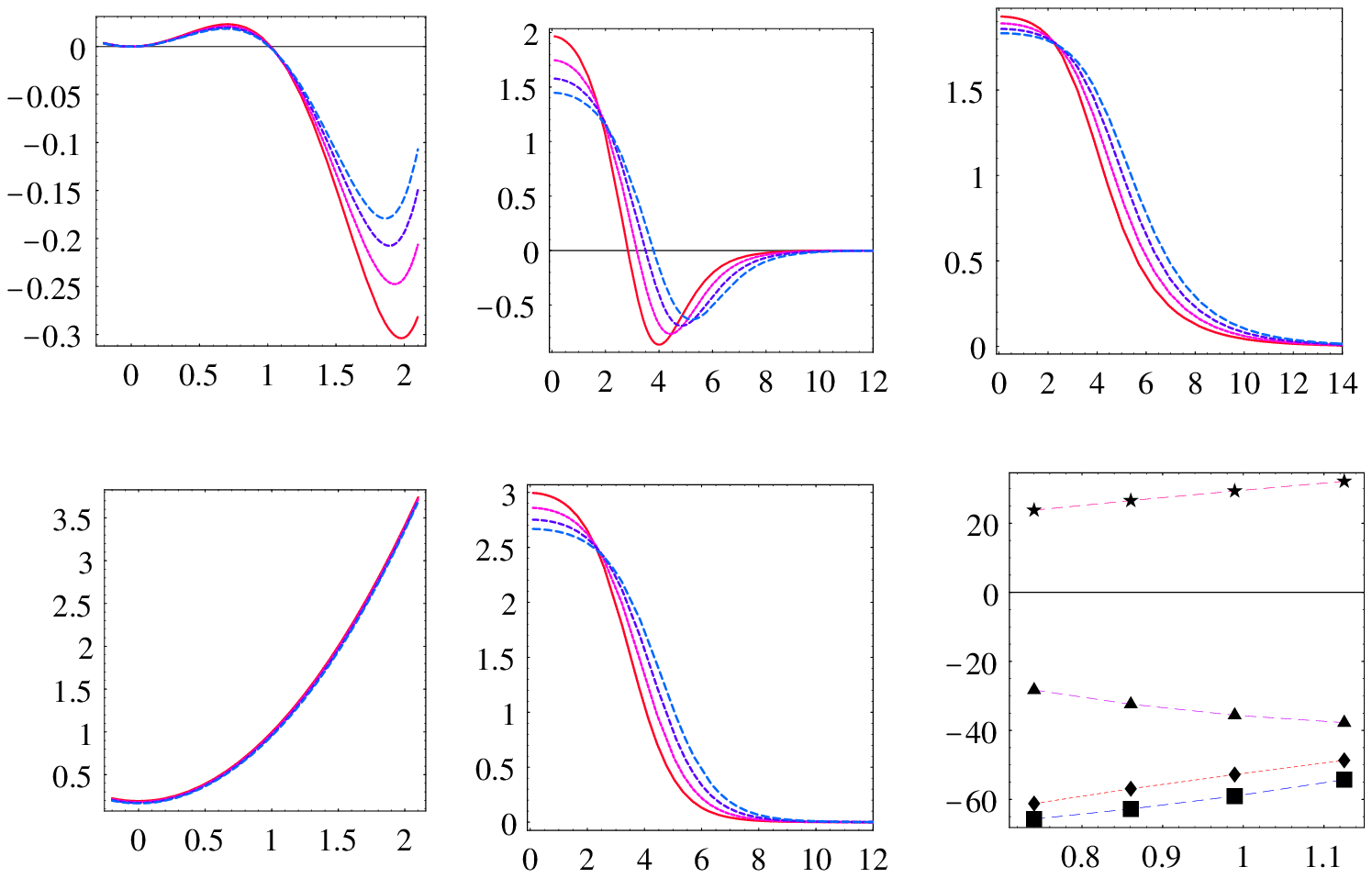}\Red
\put(3.5,10){\makebox(0,0){{\bf (\ref{fig:maxg}a)} $U_k(\phi_1,0)$}}
\put(8.4,10){\makebox(0,0){{\bf (\ref{fig:maxg}c)} $W_{1k}(r)$}}
\put(13.5,10){\makebox(0,0){{\bf (\ref{fig:maxg}e)} $\phi_{\rm b}(r)$}}
\put(3.3,4.8){\makebox(0,0){{\bf (\ref{fig:maxg}b)} $[\partial^2 U_k/\partial\phi_2^2](\phi_1,0)$}}
\put(8.4,4.8){\makebox(0,0){{\bf (\ref{fig:maxg}d) $W_{2k}(r)$} }}
\put(13.5,4.8){\makebox(0,0){{\bf (\ref{fig:maxg}f)} Results}}
\Black
\put(3.3,6.2){\footnotesize large $k$}
\put(12,1.3){\footnotesize $\ln(I/k_f^4)$}
\put(12,2.3){\footnotesize $\ln (A_{2k})$}
\put(12,3.6){\footnotesize $\ln (A_{1k}/k_f^4)$}
\end{picture}
\caption[SP]{\em As in fig.~\ref{fig:cubic}, but for a model with
initial potential given by eq.~(\ref{pot2})
with $\phi^2_{0k_0}=2~k_0$, $\lx_{k_0}=0.35~k_0$, $g_{k_0}=0.8~k_0$
and $\nu_{k_0}=0.8$.
The calculation is performed between the scales $k_i=e^{-0.1} k_0$ and $k_f=e^{-0.7} k_0$.
All dimensionful quantities are given in units of $k_f$.
\label{fig:maxg}}
\end{center}\end{figure}

\begin{figure}[t]
\begin{center}\hspace{-5mm}
\begin{picture}(16,11)
\putps(0,-0.4)(0,-0.4){fradmaxg}{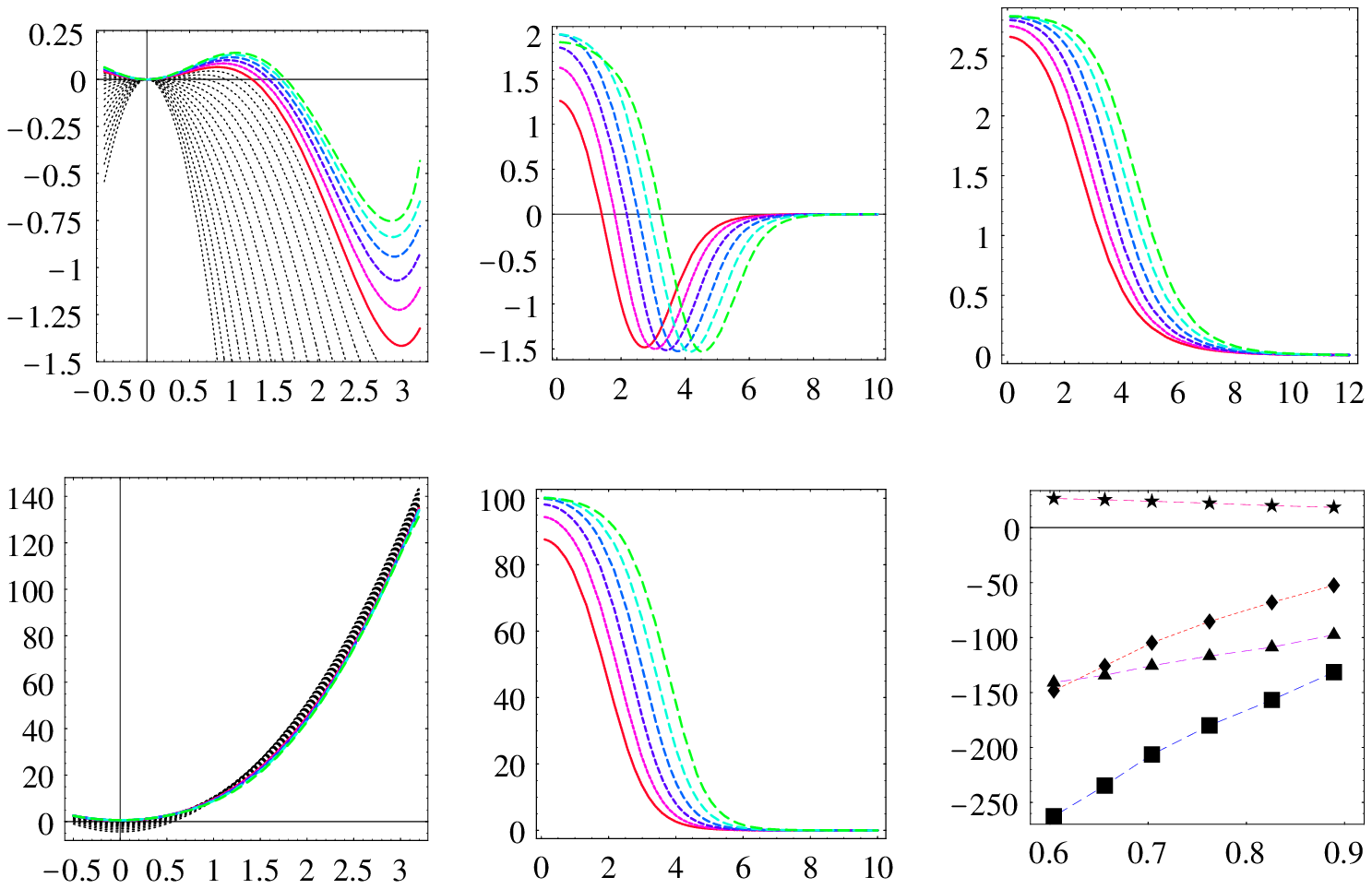}\Red
\put(3.5,10){\makebox(0,0){{\bf (\ref{fig:radmaxg}a)} $U_k(\phi_1,0)$}}
\put(8.4,10){\makebox(0,0){{\bf (\ref{fig:radmaxg}c)} $W_{1k}(r)$}}
\put(13.5,10){\makebox(0,0){{\bf (\ref{fig:radmaxg}e)} $\phi_{\rm b}(r)$}}
\put(3.3,4.8){\makebox(0,0){{\bf (\ref{fig:radmaxg}b)} $[\partial^2 U_k/\partial\phi_2^2](\phi_1,0)$}}
\put(8.4,4.8){\makebox(0,0){{\bf (\ref{fig:radmaxg}d) $W_{2k}(r)$} }}
\put(13.5,4.8){\makebox(0,0){{\bf (\ref{fig:radmaxg}f)} Results}}
\Black
\put(12,3.96){\footnotesize $\ln (A_{1k}/k_f^4)$}
\put(12,2.7){\footnotesize $-S_k$}
\put(12.3,2){\footnotesize $\ln (A_{2k})$}
\put(12.7,1){\footnotesize $\ln(I/k_f^4)$}
\end{picture}
\caption[SP]{\em As in fig.~\ref{fig:cubic}, but for a radiatively induced
first-order 
phase transition in a model with initial potential given by eq.~(\ref{pot3})
with $\phi^2_{0k_0}=1.712~k_0$, $\lx_{k_0}=0.01~k_0$ and
$g_{k_0}=0.2~k_0$.
The calculation is performed between the scales $k_i=e^{-4.7} k_0$ and $k_f=e^{-5.2} k_0$.
All dimensionful quantities are given in units of $k_f$.
\label{fig:radmaxg}}
\end{center}\end{figure}

\begin{figure}[t]
\begin{center}\hspace{-5mm}
\begin{picture}(16,11)
\putps(0,-0.4)(0,-0.4){fradming}{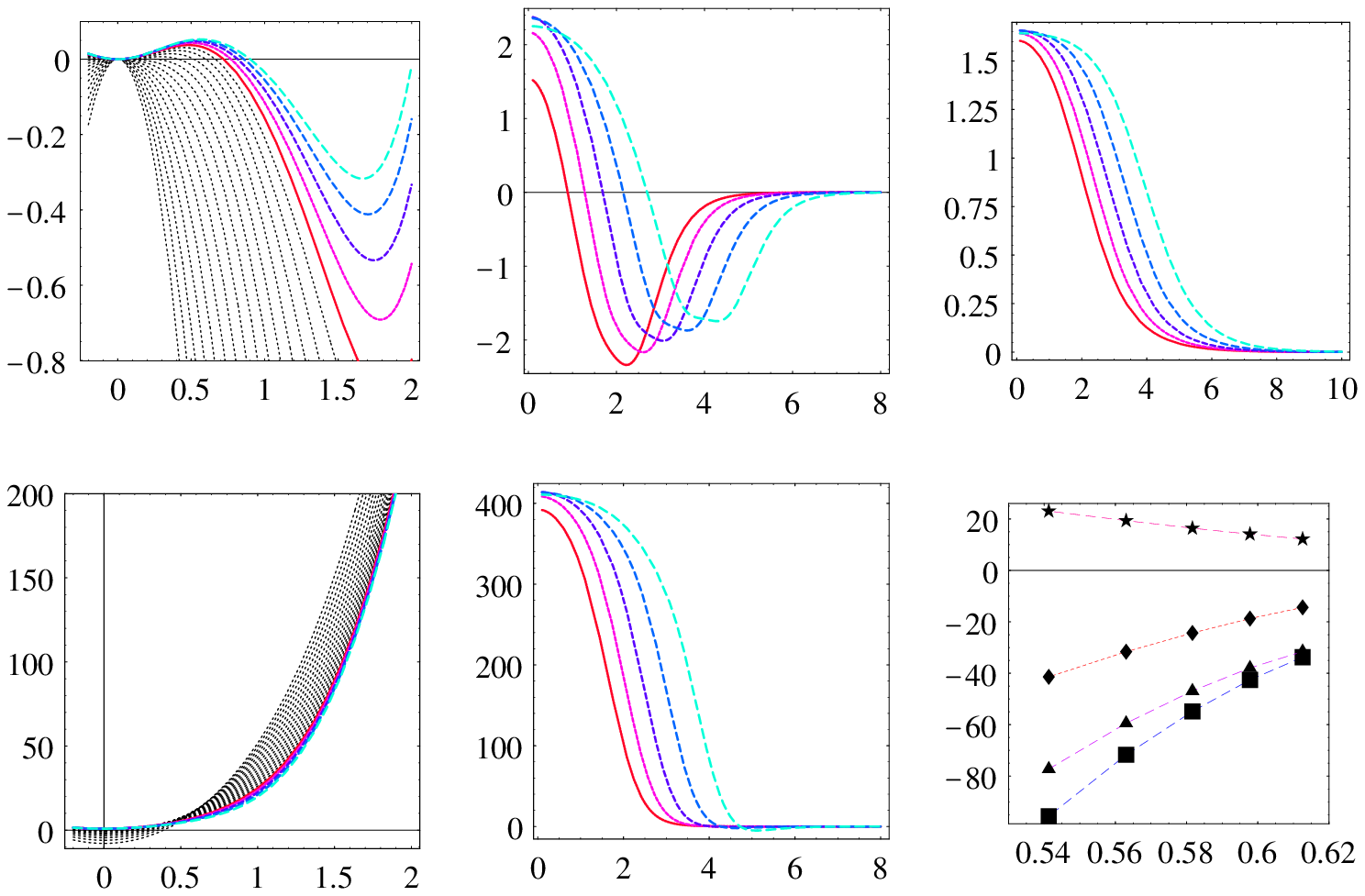}\Red
\put(3.5,10){\makebox(0,0){{\bf (\ref{fig:radming}a)} $U_k(\phi_1,0)$}}
\put(8.4,10){\makebox(0,0){{\bf (\ref{fig:radming}c)} $W_{1k}(r)$}}
\put(13.5,10){\makebox(0,0){{\bf (\ref{fig:radming}e)} $\phi_{\rm b}(r)$}}
\put(3.3,4.8){\makebox(0,0){{\bf (\ref{fig:radming}b)} $[\partial^2 U_k/\partial\phi_2^2](\phi_1,0)$}}
\put(8.4,4.8){\makebox(0,0){{\bf (\ref{fig:radming}d) $W_{2k}(r)$} }}
\put(13.5,4.8){\makebox(0,0){{\bf (\ref{fig:radming}f)} Results}}
\Black
\put(11.7,3.7){\footnotesize $\ln (A_{1k}/k_f^4)$}
\put(11.8,2.7){\footnotesize $-S_k$}
\put(11.7,2){\footnotesize $\ln (A_{2k})$}
\put(12.7,1){\footnotesize $\ln(I/k_f^4)$}
\end{picture}
\caption[SP]{\em As in fig.~\ref{fig:cubic},  but for a radiatively induced
first-order 
phase transition in a model with initial potential given by eq.~(\ref{pot3})
with $\phi^2_{0k_0}\approx0.4999~k_0$, $\lx_{k_0}=0.01~k_0$ and $g_{k_0}=0.1~k_0$.
The calculation is performed between the scales $k_i=e^{-9.2} k_0$ and $k_f=e^{-9.6} k_0$.
All dimensionful quantities are given in units of $k_f$.
\label{fig:radming}}
\end{center}\end{figure}

\setcounter{equation}{0}
\renewcommand{\theequation}{\thesection.\arabic{equation}}

\section{Results}
As a first example, we apply our formalism to a theory with a potential that
strongly resembles the ones used in refs.~\cite{first,second}.
The potential is consistent with the symmetry $\phb \leftrightarrow - \phb$ 
and near the $\pha$-axis has the form 
\be
U_{k_0}(\pha,\phb)= -J_{k_0}\pha
+\frac{1}{2} m^2_{1k_0}\pha^2 + \frac{1}{2} m^2_{2k_0}\phb^2
+ \frac{1}{8}\lx_{k_0}\left( \pha^4 + \phb^4 \right) 
+ g_{k_0} \pha^2 \phb^2.
\label{pot1} \ee
The term linear in $\pha$ can be removed through an appropriate shift
of $\pha$~\cite{seide,first}. 
This would introduce additional terms $\sim \pha^3$ 
and $\sim \pha \phb^2$. 
In fig.~1a we present the evolution of $U_k(\pha)\equiv U_k(\pha,0)$ 
for $m^2_{1k_0}=-0.1~k^2_0$, $m^2_{2k_0}=0.1~k^2_0$, 
$\lx_{k_0}=g_{k_0}=0.1~k_0$ and $J_{k_0}=0.6~k_0^{5/2}$. We always shift
the location of the false vacuum to zero. 
The evolution of $U_{22}(\pha)\equiv\partial^2U_k/\partial\phb^2(\pha,0)$
is displayed in fig.~1b. 
The solid lines correspond to $k_i/k_0=e^{-0.8}$, while the
line with longest dashes (that has the smallest barrier height)
corresponds to $k_f/k_0=e^{-1.2}$. At the scale $k_f$ the negative 
curvature at the top of the barrier is slightly larger than 
$-k_f^2$. This is the point in the evolution of the potential
where configurations that 
interpolate between the minima start becoming relevant
in the functional integral that defines the coarse-grained potential
\cite{convex,polonyi}. 
For this reason, we stop the evolution at this point. 
The potential and the field have been
normalized with respect to $k_f$, so that they are of order 1. 
The profile of the critical bubble $\phibounce(r)$
is plotted in fig.~1e in units of $k_f$
for the same sequence of scales.  For $k\approx k_f$ the characteristic 
length scale of the bubble profile and $1/k$ are comparable. This is expected, 
because the form of the profile is determined by the barrier of the potential,
whose curvature is $\approx -k^2$ at this point. 
This is an additional indication that we should not proceed to coarse-graining
scales below $k_f$.
The quantities $W_{1k}(r)=U_{11}(\phibounce(r))-U_{11}(0)$
and $W_{2k}(r)=U_{22}(\phibounce(r))-U_{22}(0)$ are plotted in
figs.~1c and 1d respectively. 

\smallskip

Our results for the nucleation rate are presented in fig.~1f.
The horizontal axis corresponds to $k/\sqrt{U_{11}(\phi_t})$,
i.e. the ratio of the scale $k$
to the square root of the positive curvature of the potential along the 
$\pha$-axis at the true vacuum. 
The latter quantity gives the mass of the field $\pha$
at the absolute minimum. 
Typically, when $k$ crosses below this mass (corresponding to
the value 1 on the horizontal axis) the massive fluctuations of the fields
start decoupling (in all the examples we present 
the mass of $\phb$ is of the same order or larger
than that of $\pha$ at the absolute minimum) 
and the evolution of the convex parts of
the  potential slows down and eventually stops.
The dark diamonds give the negative of the action $S_k$ 
of the saddle point at the scale $k$. We observe a strong 
$k$ dependence of this quantity.
The stars in fig.~1d indicate the values of 
$\ln ( A_{1k}/k^4_f )$ and the triangles those of $\ln ( A_{2k} )$,
where the two prefactors $A_{1k}$, $A_{2k}$ are defined in eqs.~(\ref{rrate}). 
Again a significant $k$ dependence is observed. More specifically,
the values of $A_{1k}$ and $A_{2k}$ 
decrease for decreasing $k$. This is expected,
because $k$ acts as the effective ultraviolet cutoff in the calculation 
of the fluctuation determinants. For smaller $k$,
fewer fluctuations with wavelengths above an increasing length scale
$\sim 1/k$ contribute explicitly to the fluctuation determinants. 
We notice that both $A_{1k}$ and $A_{2k}$ are significantly smaller than 
$S_k$ and enhance the total nucleation 
rate. The dark squares give our results for 
$\ln(I/k^4_f ) 
= -S_k+\ln  (A_{1k}A_{2k}/k^4_f )$. 
This quantity has a very small
$k$ dependence, which confirms our expectation that the 
nucleation rate should be independent of the scale $k$.
The small residual dependence on $k$ can be used to estimate the 
contribution of the next order in the expansion around the saddle point.
This contribution is expected to be smaller than the first-order correction
$\ln (A_{1k}A_{2k}/k^4_f )$. This example confirms the 
conclusions of refs.~\cite{first,second} on the region of validity of
Langer's theory of homogeneous nucleation: This
theory is applicable as long as the expansion around the semiclassical
saddle-point is convergent. 

\medskip 
In fig.~2 we consider a theory invariant under the symmetries
$\pha \leftrightarrow - \pha$, 
$\phb \leftrightarrow - \phb$, 
$\pha \leftrightarrow \phb$. 
The potential is given by 
\be
U_{k_0}(\pha,\phb)= 
 \frac{\lx_{k_0}}{8}
\left[ ( \pha^2-\phi_{0k_0}^2 )^2 
+      ( \phb^2-\phi_{0k_0}^2 )^2 \right] 
+ \frac{g_{k_0}}{4}  \pha^2 \phb^2
+ \frac{\nu_{k_0}}{48}
\left[ ( \pha^2-\phi_{0k_0}^2)^3 
+      ( \phb^2-\phi_{0k_0}^2)^3 \right],
\label{pot2} \ee
with $\phi^2_{0k_0}=2~k_0$, 
$\lx_{k_0}=0.4~k_0$, $g_{k_0}=0.3~k_0$
and $\nu_{k_0}=1$.
In fig.~2a we present the evolution of $U_k(\pha)$
and in fig.~2b that of $U_{22}(\pha)$, in complete analogy 
to figs.~1a and 1b.
The quantities $W_{1k}(r)=U_{11}(\phibounce(r))-U_{11}(0)$
and $W_{2k}(r)=U_{22}(\phibounce(r))-U_{22}(0)$ are plotted in
figs.~2c and 2d respectively. 
Our results for the nucleation rate are shown in fig.~2f.
Again, we observe a significant $k$ dependence of 
$S_k$, 
$\ln ( A_{1k}/k^4_f )$ and $\ln ( A_{2k} )$ which cancels out
in the total rate. A qualitative difference between figs.~1 and 2 
concerns the sign of $\ln ( A_{2k} )$, which is positive
in the first case and negative in the second. 
This behaviour can be understood through the comparison of the form of  
$W_{2k}(r)$ in figs.~1d and 2d. In the first case, the mass term of
the $\phb$ fluctuations in most of the interior of the critical 
bubble is smaller than that in the exterior. As a result, the operator
$-\partial^2+U_{22}(\phibounce(r))$ has lower eigenvalues
than the operator $-\partial^2+U_{22}(0)$ and the prefactor 
$A_{2k}$ in eq.~(\ref{rrate}) tends to enhance the total rate. 
The opposite is true in the second case. The $\phb$ fluctuations
become more massive in the interior of the bubble and they tend
to be suppressed. As these fluctuations are an integral part of the 
critical bubble, the total rate must be suppressed as well. 
Notice that the form of $W_{1k}(r)$ guarantees that the 
$\pha$ fluctuations always tend to enhance the nucleation rate.

In fig.~3 we present our results for
a potential given by eq.~(\ref{pot2})
with $\phi^2_{0k_0}=2~k_0$, 
$\lx_{k_0}=0.35~k_0$, $g_{k_0}=0.8~k_0$
and $\nu_{k_0}=0.8$. The behaviour is similar to that in fig.~2, but
the first-order corrections 
$\ln ( A_{1k}/k^4_f )$ and $|\ln ( A_{2k})|$ become now comparable to 
$S_k$. This signals the breakdown of the expansion around the
semiclassical saddle-point. This is confirmed by the significant 
$k$ dependence of the predicted total rate, which indicates 
that the higher-order corrections cannot be neglected anymore.
The examples of figs.~1--3 demonstrate the reliability and
consistency of our approach 
in theories with two fluctuating fields. The present work confirms 
that the picture of tunnelling we have developed is not relevant only
for the one-field case studied in refs.~\cite{first,second}.
On the contrary, it has a wide range of applicability
to many physical systems.

\medskip

We now turn to the discussion of radiatively induced first-order
phase transitions. An example can be observed in a theory defined through
the potential
\be
U_{k_0}(\pha,\phb)= 
 \frac{\lx_{k_0}}{8}
\left[ ( \pha^2-\phi_{0k_0}^2 )^2 
+      ( \phb^2-\phi_{0k_0}^2 )^2 \right] 
+ \frac{g_{k_0}}{4}  \pha^2 \phb^2,
\label{pot3} \ee
with $\phi^2_{0k_0}=1.712~k_0$, 
$\lx_{k_0}=0.01~k_0$ and $g_{k_0}=0.2~k_0$.
Our results for this theory are presented in fig.~4.
In fig.~4a we plot a large part of the evolution of 
$U_k(\pha)$. The initial potential has only one minimum along the positive
$\pha$-axis (and the equivalent ones under the 
the symmetries
$\pha \leftrightarrow - \pha$, 
$\phb \leftrightarrow - \phb$, 
$\pha \leftrightarrow \phb$) and a maximum at the origin. 
In the sequence of potentials depicted by
dotted lines we observe the appearance of a new minimum
at the origin at some point in the evolution
(at $k/k_0 \approx e^{-4.4}$). This minimum is generated by
the integration of fluctuations of the $\phb$ field, whose mass depends on
$\pha$ through the last term in eq.~(\ref{pot3})
(the Coleman-Weinberg mechanism). Detailed descriptions of this phenomenon
within the framework of the effective average action 
have been presented in refs.~\cite{bubble1,bubble2,me,
twoscalar,abelian4d,supercond} for a variety of models.
In fig. 4b it can be seen that the mass term of the 
$\phb$ field at the origin turns positive at the same value of $k$.
This is a consequence of the 
$\pha \leftrightarrow \phb$ symmetry of the potential.
We calculate the nucleation rate using the potentials of the
last stages of the evolution. 
The solid lines correspond to $k_i/k_0=e^{-4.7}$, while the
line with longest dashes 
corresponds to $k_f/k_0=e^{-5.2}$. 
In figs.~4b--4e we observe that the mass of the $\phb$ fluctuations in
the interior of the critical bubble is much larger than the other
mass scales of the problem, which are comparable to $k_f$. 
This is a consequence of our choice of couplings $g/\lx=20$. Such a large ratio
of $g/\lx$ is necessary for a strongly first-order
phase transition to be radiatively induced. Unfortunately, this range of
couplings also leads to large values for the $\phb$ mass and, as a result, to
values of 
$|\ln ( A_{2k})|$ that are comparable or larger than the
saddle-point action $S_k$, even though $\ln ( A_{1k}/k^4_f )$
remains small. As a result, the saddle-point approximation breaks down
and the predicted nucleation rate $I/k^4_f$ is strongly $k$-dependent.

It may be possible to obtain a convergent expansion around the saddle point
by considering models with smaller values of $g$. 
In fig.~5 we repeat the calculation of the nucleation rate for
a theory described by the potential of eq.~(\ref{pot3})
with $\phi^2_{0k_0} \approx 0.4999~k_0$, 
$\lx_{k_0}=0.01~k_0$ and $g_{k_0}=0.1~k_0$. Due to the smaller value of 
the ratio $g/\lx=10$, we observe a more weakly first-order phase transition.
This can be checked by considering the value of $k_f$ that sets the 
scale for the dimensionful quantities in figs.~4 and 5.
This scale is smaller by a factor $e^{-4.4}\approx 0.012$ in fig.~5 than
in fig.~4. Fig.~5f indicates that the expansion around the saddle point is
more problematic in this case. Not only
$|\ln ( A_{2k} )|$ is larger than the 
saddle-point 
action $S_k$, but the prefactor $\ln ( A_{1k}/k^4_f )$, associated with
the fluctuations of the $\pha$ field, becomes now comparable to $S_k$.
This behaviour of $A_{1k}$ for weakly first-order phase transitions 
has already been observed in refs.~\cite{first,second}.
We have not managed to find any region of the parameter space that leads
to a convergent saddle-point expansion for the nucleation rate.  
The generic behaviour resembles either the one depicted in fig.~4 for
strongly first-order phase transitions, or the one in fig.~5 for 
weakly first-order phase transitions. The implications for 
cosmological phase transitions, such as the electroweak,
are discussed in
the following section.

\setcounter{equation}{0}
\renewcommand{\theequation}{\thesection.\arabic{equation}}

\section{Discussion and conclusions}

In this paper we generalized to theories of two scalar fields
the analysis of refs.~\cite{first,second}
for the consistent description of first-order phase transitions in the 
one-field case. We established the reliability of the method by
making consistency checks for a variety of models described by 
different potentials. The conclusions of refs.~\cite{first,second}
were verified in this more general context:
Langer's theory of homogeneous nucleation is applicable 
as long as the expansion around the semiclassical
saddle point that dominates tunnelling is convergent. 
The total bubble-nucleation rate is not determined simply through the 
exponential suppression by the action of the saddle point. The 
fluctuations around the saddle point are important and generate
a pre-exponential factor which can be significant 
even in the cases for which the saddle-point expansion is convergent.
In the example of fig.~1 the logarithm of the pre-exponential factor
is about 30\% of the saddle-point action.

If the prefactor becomes comparable to the saddle-point action, the
expansion ceases to converge. In our approach this is signalled by
the appearance of a strong dependence of the predicted nucleation
rate on the coarse-graining scale at which the free energy of
the system is evaluated. We found two types of situations in
which this happens:
\begin{itemize}
\item[a)] The fluctuations of the field $\pha$, 
that serves as the order parameter
for the phase transition and varies along the saddle point, become significant
in the case of weakly first-order phase transitions. 
Typically these fluctuations enhance the total rate and can even compensate 
the exponential suppression. Several examples of this
behaviour were given in refs.~\cite{first,second}. The conclusions 
were reconfirmed by the study of the prefactor $A_{1k}$ in fig.~5. 
The reason for this behaviour can be traced to the form of the differential
operators in the prefactor associated with the field $\pha$ (see eqs. 
eqs.~(\ref{rrate})--(\ref{ak})).
This prefactor, before regularization, involves the ratio 
$\det'(-\partial^2+m^2_{1}+W_{1k}(r))/\det(-\partial^2+m^2_{1})$, with
$m^2_{1}=U_{11}(0)$ and $W_{1k}(r)=U_{11}\left(\phi_b(r)\right)
-U_{11}(0)$. The function $W_{1k}(r)$ always has a minimum away from
$r=0$ (see figs.~1--5), where it takes negative values. As a result the lowest
eigenvalues of the operator $\det'(-\partial^2+m^2_{1}+W_{1k}(r))$ are smaller
than those of $\det(-\partial^2+m^2_{1})$. The elimination of
the very large eigenvalues from the determinants
through regularization does not affect this
fact and the prefactor $A_{1k}$ is always larger than 1. Moreover,
for weakly first-order phase transitions it becomes exponentially large
because of the proliferation of low eigenvalues in 
$\det'(-\partial^2+m^2_{1}+W_{1k}(r))$. In physical terms, this implies the 
existence of a large class of field configurations of free energy comparable
to that of the saddle-point. Despite the fact that they are not 
saddle points of the free energy 
(they are rather deformations of a saddle point)
and are, therefore, unstable, they result in a dramatic increase of
the nucleation rate. This picture is very similar to that of
``subcritical bubbles'' of ref.~\cite{gleiser}. In ref.~\cite{nonpert}
the nucleation rate was
computed by first calculating a corrected potential that incorporates 
the effect of such non-perturbative configurations. The pre-exponential factor
must be assumed to be of order 1 in this approach, as the effect of most
deformations of the critical bubble has already been taken into account in
the potential.
In our approach the non-perturbative effects are incorporated
through the prefactor.
Both methods lead to similar conclusions for the enhancement of 
the total nucleation rate.
\item[b)] The fluctuations 
of the field $\phb$, that is orthogonal to $\pha$ and 
can trigger a radiatively induced first-order phase transition, can also
become important. In the radiatively induced scenario,
the mass of $\phb$ depends on the expectation value of $\pha$ through a
mixing term $\sim g \pha^2 \phb^2$ in the potential. If the coupling
$g$ is sufficiently large, a second minimum of the renormalized
potential can be generated through the integration of the $\phb$ fluctuations,
even if the bare potential has only one minimum.
In all cases where this was achieved in the models we investigated
(figs.~4 and 5 and results not presented here), we found
that the pre-exponential factor is large compared to the saddle-point action.
This is not surprising. The radiative corrections to the 
potential and the pre-exponential factor have a very similar form 
of fluctuation determinants. When the radiative corrections are large enough
to modify the bare potential and generate a new minimum, the 
pre-exponential factor should be expected to be important also. 
More precisely, the 
reason for this behaviour can be traced again to the form of the differential
operators in the prefactor.
The prefactor associated with the field $\phb$
involves the ratio 
$\det(-\partial^2+m^2_{2}+W_{2k}(r))/\det(-\partial^2+m^2_{2})$, with
$m^2_{2}=U_{22}(0)$ and $W_{2k}(r)=U_{22}\left(\phi_b(r)\right)
-U_{22}(0)$. In units in which $\phi_b(r)$ is of order 1, 
the function $W_{2k}(r)$ takes very large positive values
near $r=0$ (see figs.~4,5). This is a consequence of the
large values of $g$ that are required for
the appearance of a new minimum in the potential.
As a result, the lowest
eigenvalues of the operator $\det(-\partial^2+m^2_{2}+W_{2k}(r))$ are 
much larger 
than those of $\det(-\partial^2+m^2_{2})$. This induces a large suppression
of the nucleation rate. 
In physical terms, this implies that the deformations of the critical
bubble in the $\phb$ direction cost excessive amounts of free energy. 
As these fluctuations are inherent to the system, the total nucleation 
rate is suppressed when they are taken into account properly.
\end{itemize}

An interesting question concerns the possibility to ameliorate the
situation for the radiatively induced first-order phase transitions by
integrating out completely 
the $\phb$ field along the lines of ref.~\cite{ewein}.
One may even envisage theories in which the term $\sim \phb^4$ is absent and
the gaussian integration over $\phb$ can be peformed exactly. The 
result would be the appearance of a term 
$\sim\ln\det(-\partial^2+U_{22}(\pha))$ in the ``effective''action of the
$\pha$ field. For homogeneous field configurations, such a term would
generate the two-minimum structure in the effective potential.
However, the determination of 
the bubble profile would require the minimization of the whole ``effective''
action, including 
the additional term 
$\sim\ln\det(-\partial^2+U_{22}(\pha))$. The evaluation of the
prefactor would involve a contribution that corresponds
to a higher order in the semiclassical expansion 
that we employed in this work. The reason is that
$\ln\det(-\partial^2+U_{22}(\pha))$ 
must be expanded around the saddle point configuration $\phi_b(r)$
and the new $\pha$ integration must be performed. Thus, it is clear
that this approach does not offer any technical advantages. 
Moreover,
it is possible that an action
that includes the term 
$\sim\ln\det(-\partial^2+U_{22}(\pha))$ is problematic
due to the appearance of non-localities for field configurations for which 
$U_{22}(\pha)\approx 0$. 

Our approach is based on a separation of the problem in two steps. At first,
the high frequency modes of all fields are integrated out. The resulting 
effective theory can be approximated by a local action, as the integration 
of modes is performed with an explicit infrared cutoff so that non-localities
are suppressed. Subsequently, the nucleation problem is addressed in
a systematic expansion around the saddle point. A possibility suggested by
the discussion in the previous paragraph is that an ``improved'' saddle
point may be obtained if we minimize $S_k + \ln A_k$ instead of only
$S_k$. However, the corrections to the total rate
that are induced by this change are of the
same order as corrections from going beyond the gaussian approximation in
the expansion around the saddle point. This has been confirmed by
approximating $\ln A_k$ through eq.~(\ref{apprpref}) and minimizing 
$S_k + \ln A_k$. For $|\ln A_k|$ smaller than 
$S_k$ we have found that, although the saddle point 
is modified, the value of $S_k + \ln A_k$ remains largely
unaffected. 

Our results are relevant for most cosmological phase transitions.
For example, the electroweak phase transition for Higgs boson masses below 
$\approx 80$~GeV is a radiatively induced first-order one.
The same is true for the phase transitions in extensions of the standard
model or in grand unified theories. Usually the second minimum of the 
potential of the scalar Higgs field
appears as a result of the integration of the fluctuations of
gauge fields, whose mass is generated through the Higgs mechanism.
There is a close resemblance with the two-scalar models we considered.
This is apparent if one considers the mass term of the $\phb$ field along
the $\pha$-axis. For the potential of eq.~(\ref{pot3}), it is initially
given by $M^2_{2k_0}(\pha)=-\lx_{k_0} \phi_{0k_0}^2 /2 + g_{k_0} \pha^2/2$. 
For the range of couplings $g_{k_0} \gg \lx_{k_0}$, which is relevant for
radiatively induced first-order phase transitions, the first term in the 
previous expression is negligible. The evolution of the potential at
scales $k < k_0$ results in the replacement of $g_{k_0}$ by an effective
running coupling $g_k(\pha)$. However, in cases similar to the ones
we considered, in which no fixed points affect the evolution, the running
of $g$ is not substantial. 

In the gauged Higgs models, the mass of the
gauge fields is generated through a term $\sim e^2 \pha^2 A^{\mu} A_{\mu}$.
At the level of the bare action the mass term is
$M^2_{gk_0}(\pha)=e^2_{k_0} \pha^2$. 
Within a truncation 
that preserves the same invariants as the bare action, the evolution of
the potential is given by an equation completely analogous to
eq.~(\ref{evpot}) with $M^2_2$ replaced by $M^2_g$ (and the modification
of some numerical factors)~\cite{gauge,me,supercond,bastian}. The replacement
of $e^2_{k_0}$ by a running coupling $e^2_k(\pha)$ at scales $k<k_0$
does not modify the qualitative behaviour of the potential, 
as long as a first-order phase transition takes place~\cite{me}. 
It is very likely, therefore, that our conlcusions concerning the 
convergence of the expansion around the saddle point remain valid
in the case of the radiatively induced first-order phase transitions
in the gauged Higgs models. This would imply that the estimates
of the nucleation rate that rely only on the exponential suppression
factor may be very misleading. Moreover, the whole series may not be
convergent and this would make an alternative calculational scheme
necessary. 

A confirmation of the above possibility is provided by estimates
of the bubble-nucleation rate based on a perturbative 
approximation of the effective action. 
Such a calculation was performed in ref. 
\cite{schmidt} for the electroweak phase transition.
An effective action for the Higgs field was obtained
by integrating out the gauge fields through a perturbative calculation.
The saddle-point action was evaluated within
the effective scalar theory in the standard way, while the
pre-exponential factor was computed using the heat-kernel method. 
There are several points that need refinement in this approach: 
a) It is questionable whether the gauge fields can 
be integrated out perturbatively, as they are
massless in the symmetric phase within perturbation theory. 
b) The derivative expansion of the effective action for
the scalar fields is problematic around the origin of the
potential, where the gauge theory becomes strongly coupled. 
c) In the absence of a coarse-graining scale the fluctuations
of the scalar fields cannot be treated properly, as the problem
of double-counting their effect in the effective action and
the prefactor is not resolved. 
It is remarkable, however, that, despite the above problems,
the conclusions of ref.~\cite{schmidt} are in 
agreement with the implications of our results: 
The pre-exponential factor is significant and results in
the additional suppression of the total rate. Moreover, the saddle-point
approximation is problematic near the critical temperature. 

\medskip

The application of our formalism to the electroweak phase transition
must deal with several technical points. Firstly, an efficient truncation of
the average effective action must be developed, appropriate for
the discussion of gauged Higgs theories. Such truncations have been
discussed in refs.~\cite{gauge,me,supercond,bastian} and
the resulting evolution equations for the potential and the gauge coupling
have been obtained. In the case of the $SU(2)$ Higgs model, they account
for the growth of the gauge coupling near the origin of the potential
and the emergence of a strongly coupled regime. They also take into account 
the contribution of the Higgs-field fluctuations.
The saddle-point action can be evaluated along the lines of 
refs.~\cite{bubble1,bubble2,first} and the present work.
The calculation of the pre-exponential factor is more difficult. 
The fluctuation determinants associated with the gauge
fields must be computed
in complete analogy to those of the $\phb$ field in this work. 
This means that they must be evaluated in the background of the saddle point,
with an ultraviolet regularization that matches the cutoff procedure in
the derivation of the evolution equations for the potential and
the gauge couplings. Particular care 
is required for the treatment of additional zero eigenvalues that result from
the presence of Goldstone modes in the phase with symmetry breaking 
\cite{buch}. However, we believe that the above technical points should
not affect the qualitative conclusion reached in this work: If the radiative
corrections can affect the vacuum structure of the theory, the pre-exponential
factor in the tunnelling rate is significant.

Another possible application concerns the bound on the Higgs-boson mass 
from the requirement that our vacuum is not destabilized by the presence 
of a non-standard 
vacuum at large Higgs-field values~\cite{stab,jose}. 
This new vacuum is generated
by the top-quark radiative corrections to the effective potential. The most
stringent bound arises from the requirement that
the probability of thermal tunnelling 
from the symmetric phase to the non-standard vacuum at
some temperature above the critical temperature of the
electroweak phase transition is negligible.
This problem can be treated within an effective three-dimensional
theory at energy scales below the temperature. At such scales the top quark 
can be integrated out, as its propagator 
contains an effective mass of the order of the
temperature. Its effect can be incorporated in the initial conditions
for the evolution equations for the potential and the gauge couplings.
The potentials that are relevant for this problem have a form
very close to that along the spinodal line (the false vacuum
is at the end of the metastability range 
and very close to becoming unstable)~\cite{jose}. 
It is, therefore, conceivable that the prefactor generated
by the Higgs-field fluctuations is significant~\cite{second}. The 
effect of the gauge-field fluctuations must be taken into account as well,
in analogy to the calculation of the nucleation rate for
the electroweak phase transition.

\paragraph{Acknowledgements} We  would like to thank
R. Barbieri, J. Espinosa and C. Wetterich for helpful discussions.
The work of N.T. was supported by the E.C. under TMR contract 
No. ERBFMRX--CT96--0090.

\appendix
\setcounter{equation}{0}
\renewcommand{\theequation}{\thesection.\arabic{equation}}


\end{document}